# A Mathematical Model for the Genetic Code(s) Based on Fibonacci Numbers and their q-Analogues

Tidjani Négadi

Physics Department, Faculty of Science,
University of Oran, Es-Sénia, 31100, Oran, Algeria
Email : tnegadi@gmail.com

**Abstract.** This work aims at showing the relevance and the applications possibilities of the Fibonacci sequence, and also its q-deformed or "quantum" extension, in the study of the genetic code(s). First, after the presentation of a new formula, an indexed double Fibonacci sequence, comprising the first six Fibonacci numbers, is shown to describe the 20 amino acids multiplets and their degeneracy as well as a characteristic pattern for the 61 meaningful codons. Next, the twenty amino acids, classified according to their increasing atom-number (carbon, nitrogen, oxygen and sulfur), exhibit several Fibonacci sequence patterns. Several mathematical relations are given, describing various atom-number patterns. Finally, a q-Fibonacci simple phenomenological model, with q a real deformation parameter, is used to describe, in a unified way, not only the standard genetic code, when q=1, but also all known slight variations of this latter, when q~1, as well as the case of the 21st amino acid (Selenocysteine) and the 22nd one (Pyrrolysine), also when q~1. As a by-product of this elementary model, we also show that, in the limit q=0, the number of amino acids reaches the value 6, in good agreement with old and still persistent claims stating that life, in its early development, could have used only a small number of amino acids.

**1. Introduction**
The aim of this work is to show the relevance, and also the usefulness, of the Fibonacci sequence in the study of the genetic code mathematical and chemical structure. At the very beginning of the thirteenth century, Leonardo Bonacci, better known as Fibonacci, wrote a book, *Liber Abaci,* in which he presented, among others, his famous rabbit problem the solution of which is known as the Fibonacci sequence. Eight hundred years later, in the present twenty-first century, this sequence, and the related golden section (or golden ratio), still play an important role in many areas of science. Mathematically, this sequence:

$$0, 1, 1, 2, 3, 5, 8, 13, 21, 34, 55, 89, 144, 233, \ldots \qquad (1)$$

is given by the recurrence relation $F_{n+1}=F_n+F_{n-1}$ (n=1, 2, 3, ...) with the initial conditions $F_0=0$, $F_1=1$. This sequence is known to occur in nature, in arts, and in many areas of science. In physics, there are many examples and we shall select only some few of them, for illustration. The so-called Fibonacci anyons are promising objects in the topological quantum computing research (Vaesi and Barkeshli, 2014). Here, the Fibonacci numbers are directly connected to the dimensions of the topological Hilbert spaces in which the computations are supposed to take place. The Fibonacci sequence and the golden section are at the root of the study of quasicrystals, which have been discovered in 1984 and eventually led their discoverer Dan Schechtman to win the Nobel Prize in 2011 (Shechtman et al., 1984). Still in the quantum world, it is worth mentioning the recent discovery, in 2010 (Coldea et al., 2010), and for the first time, by a team of researchers from Germany and Great Britain, of a nanoscale symmetry, hidden in solid-state matter with the golden ratio as signature. In chaos theory, (Linage et al., 2006), it has been shown that the Fibonacci sequence appears within the "Feigenbaum scaling of the period-doubling cascade to chaos" which has been observed in many dynamical systems (turbulence, cell biology, chemical oscillators). Still in physics, a quick search at the Cornell University e-prints repository (arXiv.org) shows more than 300 titles, all bearing the name Fibonacci in their titles. In biology, there are also many examples. The best known one is in Phyllotaxis (Mitchison, 1977), a branch of plant morphology, studying the regular patterns as leaves around a stem, scales on a pine cone, and so on. In a great majority of the cases, the Fibonacci numbers are involved. Another example is in the development and structure of animal forms (Wille, 2012). Also, an interesting connection between nucleotide frequencies in human single stranded DNA and Fibonacci numbers has been reported (Beleza Yamagishi and Shimabukuro, 2008). Rakočević (1998), has shown that "the Golden Mean is a key determinant of the genetic code". Still in the context of the genetic code, Gonzalez (2008) used Fibonacci's non-power redundant number representation systems to study its degeneracy structure. Recently (Négadi, 2014), we have taken the number of atoms in the four RNA ribonucleotides UMP, CMP, AMP and GMP, 144, a distinguished Fibonacci number ($144=F_{12}=12^2$), as a key parameter. This number is interesting because it is the only square Fibonacci number, besides the trivial cases of 0 and 1; this result has been established by several mathematicians (Cohn, 1964). It is also equal to the square of its index. From this latter, we derived, among others, a mathematical relation describing the "*condensation*" of the three sub-units a nucleobase, a ribose and a phosphate group, to form a ribonucleotide, and also the atom-number content of the 64 RNA-codons, according to Rakočević's pattern "2×3456", and obeying an "*invariance*" equation. We derived also the nucleon content of the amino acids in various degeneracy patterns, as well as the multiplet structure of the genetic code, using selected sums of Fibonacci numbers. In the present work, we shall continue our investigation on the relevance, and also the usefulness, of the Fibonacci sequence in the study of the genetic code. *We emphasize that the Fibonacci sequence, here, is taken as a mathematical model, just as a mathematical function could be used as a model* (see section 5). In section 2, we begin by presenting, for the first time, a new (compact) formula, relying on the use of geometric and arithmetic progressions, and giving the multiplets of the standard genetic code: 5 quartets, 3 sextets, 9 doublets, 1 triplet and 2 singlets. Next, we consider an indexed double sequence comprising the first six Fibonacci numbers and derive from it, among others, the number 61 of meaningful codons, partitioned into 20 amino acids and 41 degenerate codons. We also establish, from this double sequence, a codon-pattern characterizing (i) the division of the genetic code table into two equal moieties (of 10 amino acids each), one having pyrimidines, as the first base, and the other having purines, as the first base, and (ii) another division of the genetic code table into two moieties (of 10 amino acids each), one classifying the amino acids belonging to class-I Aminoacyl t-RNA Synthetases and the other classifying the amino acids belonging to class-II Aminoacyl t-RNA Synthetases. In section 3, we consider a classification of the 20 amino acids according to their increasing atom-number, *carbon, oxygen, nitrogen and sulfur* (CNOS), following an idea by Trifonov, and show that it exhibits several Fibonacci sequences, when split into two distinct sub-sets. Using the indexed double Fibonacci sequence, of section 2, or part of it, we derive the detailed chemical composition of the amino acids, in the above classification. We also derive the detailed atom-numbers for the above mentioned sub-sets, from the carbon atom-numbers only which, we show, exhibits also a multiplicity Fibonacci sequence. Other interesting results are derived as, for example, the missing hydrogen atom content and also the nucleon content in the above classification. In section 4, we consider a q-deformation model, inspired from mathematical physics technics, and relying on the use of q-Fibonacci numbers. Depending on the values of the phenomenological parameter q, we establish relations unifying (i) the standard genetic code (q=1), (ii) all its known variants (q ~1) and (iii) the genetic code with one or two additional amino acids such as Selenocysteine and Pyrrolysine (q~1). It is also shown that the special value q=0 of the deformation parameter leads to a small number of amino acids, in good agreement with predictions made by several existing theories on the origin of the genetic code. We end this paper by giving a brief summary and some remarks, in section 5.

## 2. The Fibonacci numbers and the multiplet structure of the standard genetic code

The genetic code, that is, the "dictionary" which ensures the translation, in the ribosome, of the genetic information contained in the genes to proteins, has been (experimentally) deciphered fifty years ago (Nirenberg et al., 1965). This discovery has been a monumental achievement in the biological sciences. Not to mention a great number of talented scientists who made earlier decisive advances, the leading figure is, without any doubt, Marshall Nirenberg, a great scientist of the twentieth century. He received the Nobel Prize in 1968, together with Robert Holley and Har Gobind Khorana, "for their interpretation of the genetic code and its function in protein synthesis". Marshall Nirenberg died in January 2010. For an historical account, one could also read the interesting hereafter cited paper, by Erdmann and Barciszewski, written at the occasion of the "50th Anniversary of the Discovery of the Genetic Code" (Erdmann and Barciszewski, 2011), where the authors summarize the achievements, and the key experiments, that led to its discovery.

Before proceeding to the main subject of this section, let us recall, briefly, some basic facts about the (standard) genetic code mathematical structure and present also, for the first time, a new formula. The genetic code is a mapping from 64 codons (triplets of nucleotides) to 20 amino acids and 3 stop codons. This mapping is many-to-one: more than one codon may correspond to an amino acid; this phenomenon is called degeneracy. The genetic code table (Table 2) is organized into group-numbers {1, 2, 3, 5, 9} and class-numbers {3, 1, 6, 4, 2}. The group-number is the number of amino acids within a given degeneracy class, labeled by its class-number; this latter being the number of codons which code an amino acid within the class. For example the group-number 9 corresponds to the class-number 2, i.e., 9 doublets. To date, to the best of our knowledge, there exist no mathematical formula to compute exactly all these numbers. Some forty years ago, it was pointed out by Gavaudan (1971) that the group-numbers, corresponding to the even-classes, are in accordance with a geometrical progression, $2^n+1$ (n=1, 2, 3), when the group-numbers are inversely ordered by an arithmetical progression (6, 4, 2). No formula was given for these latter even class-numbers. Here, we could exploit this interesting observation to go farther and derive a complete formula, including also the group-numbers of the odd class-numbers. As a matter of fact, the following expression unifies all the cases:

$$\sum_{K,k} \left(2^k + \delta_{K,0}\right)\left(8 - 2k - 5\delta_{K,1}\right) \tag{2}$$

In this formula, $\delta_{i,j}$ are Kronecker delta functions equal to 1 for i=j and 0 for i≠j. In the first parenthesis, $2^k+\delta_{K,0}$, gives the group-numbers and, in the second one, $8-2k-5\delta_{K,1}$, gives the class-numbers. For K=0 and k=1, 2, 3 we have the group-numbers 3, 5 and 9 for the even class-numbers 6, 4 and 2, respectively, and for K=1 and k=0, 1 we have the group-numbers 1 and 2 for the odd class-numbers 3 and 1, respectively. In the detail, we have

$$\sum_{k=1,2,3} \left(2^k + 1\right)(8 - 2k) + \sum_{k=0,1} 2^k (3 - 2k) \tag{3}$$
$$= (3 \times 6 + 5 \times 4 + 9 \times 2) + (1 \times 3 + 2 \times 1) = (18 + 20 + 18) + (3 + 2) = 61$$

in perfect agreement with the well-known experimental data for the standard genetic code: 3 sextets (18 codons), 5 quartets (20 codons), 9 doublets (18 codons), 1 triplet (3 codons) and 2 singlets (1 codon each). These latter (sextets, quartets, …) are also called *multiplets*. Let us remark that by subtracting Eq.(2) (or Eq.(3)) from 64 (=$2^6$), the total number of codons, we get 3 the number of stop codons or, by rearranging, 61+3=64. Now, we turn to the main subject of this section. Recently (Négadi, 2014), we have used a small portion of the Fibonacci sequence, in fact a double sequence, comprising the first six Fibonacci numbers, to show that it could house, at the same time, the number of amino acids and the degeneracy. In this section, we re-reconsider this *indexed double sequence* (Table 1) where i (=1, 2, …, 6) is the index of the Fibonacci numbers $f_i$ and $F_i$ which are, as it is shown below, respectively connected to the degeneracy and to the number of amino acids. As a matter of fact, consider the following sum

$$\sum_{1}^{6} F_i + (i + f_i) = 20 + 41 = 61 \tag{4}$$

**Table 1.** The first six Fibonacci numbers

| i    | 0 | 1 | 2 | 3 | 4 | 5 | 6 |
|------|---|---|---|---|---|---|---|
| $f_i$ | 0 | 1 | 1 | 2 | 3 | 5 | 8 |
| $F_i$ | 0 | 1 | 1 | 2 | 3 | 5 | 8 |

The first number, 20, the sum of the six Fibonacci numbers $F_i$ is equal to the number of amino acids. The second, 41, as the sum of the six Fibonacci numbers $f_i$ and their indices is equal to the total number of degenerate codons. Here also, as above, we could have, instead of just taking the sum in Eq.(4), computed the difference between $64=4^3$, the total number of codons, and the said sum. In this case, we would get 64-(20+41)=3, that is, the number of stop codons and, rearranging we end up with 20+41+3=61+3=64. In so doing, we get, at the same time, the number of amino acids, the number of degenerate codons and the number of stops. This is the same result as the one obtained from Eq.(2) above. We could also find, from the above table, the group-numbers of the genetic code as follows. First, we take the sum of the six Fibonacci numbers $F_i$, as in the above sum in Eq.(4), and write it as

$$F_1 + F_3 + F_4 + F_5 + (F_2 + F_6) = 1 + 2 + 3 + 5 + 9 = 20 \qquad (5)$$

These are, exactly, the group-numbers of the multiplets, that is, the number of amino acids in a given degeneracy class (see above). We have therefore 1 triplet, 2 singlets, 3 sextets, 5 quartets and 9 doublets, as it must be. Note that, for the doublets, we have written their number, 9, as $F_2+F_6=1+8$ or $1+2^3$, in agreement with Eq.(3) for K=0 and k=3. This writing is also interesting because of the following. We know that the sum of the indices of the six Fibonacci numbers $F_1$, $F_2$, ..., $F_6$ is equal to 21, itself a Fibonacci number. Now, the sum of the indices of $F_2$ and $F_6$ is equal to 2+6=8 and the sum of the indices of the rest of the Fibonacci numbers is equal to 13 (=1+3+4+5) so that we have 13+8=21. This is just the recurrence relation defining the Fibonacci number 21: $F_8=F_7+F_6$. These two facts seem to strengthen (or justify) the above choice for the doublets. Second, we can sort the twelve numbers constituting the Fibonacci numbers $f_i$ and their indices as follows

| multiplets | degeneracy |
|---|---|
| doublets: $F_2+F_6=1+8=9$ | $f_2+f_6=9$ |
| quartets: $F_5=5$ | $5+f_3+f_4+f_5=15$ |
| triplet: $F_1=1$ | $f_1+1=2$ |
| singlets: $F_3=2$ | 0 |
| sextets: $F_4=3$ | 2+3+4+6=15 |
| total: 20 | total: 41 |

Let us note that, for the sextets, the choice 2+3+4+6=15 is in agreement with the degeneracies (see Table 2): 2 for the two degenerate doublets of Leucine ($L^{II}$), 3 for the three degenerate codons of Arginine ($R^{IV}$), 4 for the two degenerate doublets of Arginine ($R^{II}$) and serine ($S^{II}$) and finally 6 for the six degenerate codons of Serine ($S^{IV}$) and Leucine ($L^{IV}$). It is also possible to derive from table 1 another interesting pattern, in two ways. Let us consider, first, the *even* Fibonacci numbers $F_i$ and $f_i$, on the one hand (with their indices not necessarily even), and the *odd* ones, on the other (with their indices not necessarily odd). We have

$$\begin{aligned}(F_3 + F_6) + (f_3 + f_6 + 3 + 6) &= 10 + 19 = 29 \\ (F_1 + F_2 + F_4 + F_5) + (f_1 + f_2 + f_4 + f_5 + 1 + 2 + 4 + 5) &= 10 + 22 = 32\end{aligned} \qquad (6)$$

with, of course, total sum 61. It is also possible to get this pattern in the following other manner. We consider now the even Fibonacci numbers $F_i$ and $f_i$ and the even indices, on the one hand, and the odd Fibonacci numbers $F_i$ and $f_i$ and the odd indices, on the other. We have

$$(F_3 + F_6) + (f_3 + f_6 + 2 + 4 + 6) = 10 + 22 = 32$$
$$(F_1 + F_2 + F_4 + F_5) + (f_1 + f_2 + f_4 + f_5 + 1 + 3 + 5) = 10 + 19 = 29 \quad (7)$$

This is the same pattern as above. This pattern (10+22=32, 10+19=29) shows itself in two interesting cases. In the first case, the upper half of table 2, where the codons have U or C (pyrimidines) as the first base, has 10 amino acids and 19 degenerate codons and the lower half of the table, where the codons have A or G (purines) as the first base has also 10 amino acids and 22 degenerate codons. The second case concerns the classification of the 20 Aminoacyl t-RNA Synthetases (AARRs) which are known to establish the genetic code, biochemically. These are divided into two classes of 10 members in each:

- Class-I AARRs: Y(2), W(1), C(2), Q(2), L(6), R(6), M(1), I(3), E(2), V(4)

- Class-II AARRs: F(2), N(2), H(2), K(2), D(2), S(6), P(4), T(4), G(4), A(4)

Here, we have represented the various AARRs by their corresponding amino acid and, in the parenthesis, we gave the number of codons which code it. It is not difficult to see that there are 19 degenerate codons in Class-I and 22 degenerate codons in Class-II, in nicely agreement with the above pattern. (Recall that the number of degenerate codons for a given class-number is equal to the class-number minus one.)

**Table 2.** The (standard) genetic code table.

| UUU F | UUC F | UCU S | UCC S | CUU L | CUC L | CCU P | CCC P |
|---|---|---|---|---|---|---|---|
| UUA L | UUG L | UCA S | UCG S | CUA L | CUG L | CCA P | CCG P |
| UAU Y | UAC Y | UGU C | UGC C | CAU H | CAC H | CGU R | CGC R |
| UAA stop | UAG stop | UGA stop | UGG W | CAA Q | CAG Q | CGA R | CGG R |
| AUU I | AUC I | ACU T | ACC T | GUU V | GUC V | GCU A | GCC A |
| AUA I | AUG M | ACA T | ACG T | GUA V | GUG V | GCA A | GCG A |
| AAU N | AAC N | AGU S | AGC S | GAU D | GAC D | GGU G | GGC G |
| AAA K | AAG K | AGA R | AGG R | GAA E | GAG E | GGA G | GGG G |

## 3. The Fibonacci sequence in the 20 amino acids of the genetic code
Trifonov (2001), in a presentation on the chronology and the evolution of the amino acid "alphabet", arranged the 20 amino acids according to their increasing atom-numbers, considering carbon, nitrogen, oxygen and sulfur (not hydrogen). We shall return to this interesting arrangement in the last section. Here, we follow this idea and construct, accordingly, the table below (Table 3).

**Table 3.** The 20 canonical amino acids, sorted according to their increasing C/N/O/S atom-numbers.

| #aas | AA | #Atoms | #NOS | #aas | AA | #Atoms | #NOS | #NOS |
|---|---|---|---|---|---|---|---|---|
| 1 | Gly* | 0 | 0 | 3 | Gln | 5 | 2 | 5 |
| 1 | Ala* | 1 | 0 | | Glu | 5 | 2 | |
| 2 | Ser* | 2 | 1 | | Lys | 5 | 1 | |
| | Cys | 2 | 1 | 2 | Arg | 7 | 3 | 3 |
| 3 | Pro* | 3 | 0 | | Phe | 7 | 0 | |
| | Thr* | 3 | 1 | 1 | His | 6 | 2 | 2 |
| | Val* | 3 | 0 | 1 | Tyr | 8 | 1 | 1 |
| 5 | Asn | 4 | 2 | 1 | Trp | 10 | 1 | 1 |
| | Asp* | 4 | 2 | | | $S^{II}$ | | |
| | Ile | 4 | 0 | | | | | |
| | Leu | 4 | 0 | | | | | |
| | Met | 4 | 1 | | | | | |

$S^{I}$

In this table, there are two distinct sets $S^I$ and $S^{II}$. Moreover the second set, $S^{II}$, is split, by ourselves, into two additional sub-sets, an upper part with prime atom-numbers (5 and 7) and a lower one with non-prime atom-numbers (6, 8 and 10), separated by a double line. This disposition is interesting as it will be shown below. In the second columns (AA) the amino acids are indicated, in the three-letter code. The third columns give the number of atoms Carbon, Nitrogen, Oxygen and Sulfur (CNOS) while only nitrogen, oxygen and sulfur atoms (NOS) are given in the fourth columns. As for the first columns, they give the number of amino acids for a given number of atoms according to the third columns. Finally, the fifth column, in $S^{II}$, gives the number of NOS-atoms corresponding to the sets of amino acids in the first column. Note that all quantities used in this paper (atom numbers, carbon atom numbers and nucleons numbers) correspond to the side-chains of the amino acids (see footnote 1). Let us examine this table in the detail. What is immediately seen, in $S^I$, is that the numbers in the first columns (#aas) are the first members of the Fibonacci sequence (see below). We have 1, 1, 2, 3, 5, as the number of amino acids corresponding to the five sub-sets with 0, 1, 2, 3 and 4 atoms (CNOS), respectively. In $S^{II}$, the *same* Fibonacci sequence appears for, this time, the number of NOS-atoms corresponding to amino acids sets with 10, 8, 6, 7 and 5 CNOS-atoms, respectively (see the fifth column of $S^{II}$). Along with the latter *ascending* sequence, but this time for the CNOS-atoms, we have 1, 1, 1, 2 and 3 amino acids, respectively. Here, it seems that we have something different from a Fibonacci sequence (because of the three consecutive ones) but, in fact, it is related to the Fibonacci sequence; let us see how. The Fibonacci sequence is also known to be an example of a 1D quasiperiodic structure known as the Fibonacci chain. It could be constructed from two elements S (Small) and L (Large), using the substitution rule L→LS and S→L. The first few terms are depicted in Table 4. In the third and fourth columns the number of Ls and Ss are shown and the first column gives their sum. The sequence of the Ls is the usual Fibonacci sequence {0, 1, 1, 3, 3, 5, …}, with initial conditions $F_0=0$ and $F_1=1$, the sequence of the Ss is another Fibonacci sequence {1, 0, 1, 1, 2, 3, 5, …}, with initial conditions $F_0=1$ and $F_1=0$, and their sum gives also a Fibonacci sequence (1, 1, 2, 3, 5, 8, …). There are two other known simple practical constructions of the Fibonacci sequence, the original (historical) Fibonacci's one, the solution of the "rabbit problem" (Knott, 2010) and also the "genealogical tree of a male bee" (Basin, 1963). In both these constructions, one finds the same patterns as the ones for #L and #S in Table 4. In the rabbit problem, these are the sequences of the number of pairs of "mature" rabbits and the number of pairs of "immature" rabbits, respectively. In the genealogical tree of male bees' case, we have the sequences of the "female" ancestors and the sequence of the "male" ancestors, respectively. In Table 4, corresponding to these, we have the sequence of Ls and the sequence of Ss, respectively. The connection with what was said above concerning the ascending sequence of CNOS atoms {1, 1, 1, 2, 3} is clear: except for a missing "0" at the second position, this is essentially the same as the sequence {1, 0, 1, 1, 2, 3, …} mentioned above and we could imagine that this missing "0" corresponds to 9, the (sole) missing number of CNOSs. Note also that this last sequence is essentially the same as the sequence of the differences $F_{k+1}-F_k$ (k=0, 1, 2, …).

**Table 4**: The Fibonacci chain

| Fibonacci number | Fibonacci sequence | #L | #S |
|---|---|---|---|
| 1 | S | 0 | 1 |
| 1 | L | 1 | 0 |
| 2 | LS | 1 | 1 |
| 3 | LSL | 2 | 1 |
| 5 | LSLLS | 3 | 2 |
| 8 | LSLLSLSL | 5 | 3 |
| 13 | LSLLSLSLLSLLS | 8 | 5 |
| 21 | LSLLSLSLLSLLSLSLLSLSL | 13 | 8 |
| … | ……………………… | … | … |

Now, returning to $S^I$, we observe that for a given Fibonacci number $F_i$ the corresponding number of atoms is just i-1 so that the total number of atoms in this part is given by the following sum

$$\sum_{i=1}^{5}(i-1) \times F_i = 0 + 1 + 4 + 9 + 20 = 34 \tag{8}$$

Interestingly, it appears that this sum is also a Fibonacci number, $F_9$=34. The importance of this last number will be emphasized at the end of this section in connection with to the (missing) number of hydrogen atoms from Table 3. It is well known that the Fibonacci numbers are connected to the binomial coefficients through Lucas's formula (1876):

$$F_n = \sum_{k=0}^{\lfloor (n-1)/2 \rfloor} \binom{n-k-1}{k} \tag{9}$$

where $\binom{n-k-1}{k}$ is a binomial coefficient and $\lfloor x \rfloor$ is the "floor" function which means the greatest integer less than or equal to x. Applying the above formula to the Fibonacci number $F_9$= 34 yields

$$34 = \binom{8}{0} + \binom{7}{1} + \binom{6}{2} + \binom{5}{3} + \binom{4}{4} = 1 + 7 + 15 + 10 + 1 = 10 + 24 \qquad (10)$$

where we have, in the last step, separated the odd binomials from the even binomial, 10, to get 34=10+24. This is interesting because we have 10 atoms for the odd atom-numbers (1+3×3) and 24 atoms for the even atom-numbers (2×2+5×4). This last pattern of atom-numbers in $S^I$ could also be computed from Table 1 by using Euler's phi-function φ. The φ-function of a positive integer n greater than 1 is defined to be the number of positive integers less than n that are coprime to n, with φ(1)=1; it is easily computable using Maple, for example. This function has been recently very useful in the study of the genetic code (Négadi, 2014). Consider the following sum, where we use the φ-function of the numbers of the sequence $F_i$ (from Table 1) and also their indices

$$\sum_{i=1}^{6} i + \varphi(F_i) = \sum_{i=1}^{5} i + \varphi(F_i) + (6 + \varphi(F_6)) = 24 + 10 \qquad (11)$$

By splitting the above sum into two parts, we find the same pattern found in Eq.(10). Note that the sum of the i's, on the one hand, gives 21 and the sum of the $\varphi(F_i)$'s, on the other, gives 13 and 21+13=34, i.e., the ninth Fibonacci number as defined by the recurrence relation $F_9$= $F_8$+$F_7$. Now, we turn to the second part of Table 3. The total number of atoms, here, is equal to 53 (not very far from the tenth Fibonacci number 55). There are at least two ways to compute the number of atoms in this case. One of them consists in computing the following sum involving the Fibonacci numbers $F_i$ from Table 1 and their φ-functions

$$\sum_{i=1}^{6} 2F_i + \varphi(F_i) = 53 \qquad (12)$$

We see therefore that it gives the correct number of (CNOS) atoms 53. By splitting this sum into the following two parts

$$2F_6 + \varphi(F_6) + \sum_{i=1}^{5} \varphi(F_i) = 29, \qquad \sum_{i=1}^{5} 2F_i = 24 \qquad (13)$$

we get the 53=29+24. This is the atom-number pattern in the two sub-parts of $S^{II}$: 29 atoms for the prime atom-numbers (3×5+2×7) and 24 atoms for the non-prime atom-numbers (6+8+10). Splitting the above sum, another way, gives

$$\sum_{i=1}^{4} 2F_i + \varphi(F_i) + \sum_{i=5}^{6} 2F_i + \varphi(F_i) = 19 + 34 = 53 \qquad (14)$$

This pattern, 19+34=53, is also interesting (see below). The second sum (for i=5, 6) could also be split to give

$$\sum_{i=5}^{6} 2F_i = 26, \quad \sum_{i=5}^{6} \varphi(F_i) = 8 \tag{15}$$

which gives the partition of the number of atoms in $S^I$, 34, into the carbon atom-number, 26, on the one hand, and the NOS atom-number, 8, on the other. The carbon atom-number is obtained as the number of atoms minus the number of NOS atoms (see Table 3). The sum 19+34=53, which gives the total number of atoms in $S^{II}$, is identical, in form, with the result of computing it from the number of atoms in $S^I$, *only*. To see how, we use the $a_o$-function which gives the sum of the prime factors of a number from its prime factorization (see Négadi, 2007 and Négadi, 2009 for many applications). As the factorization of the number 34 is 2×17, we have 34+$a_o$(34)=34+19=53. Finally, the sum in Eq.(13) giving 29, the number of atoms for the prime atom-numbers, could also be written as

$$F_6 + \varphi(F_6) + \sum_{1,2,3} \varphi(F_i) = 15, \quad F_6 + \varphi(F_6) + \sum_{4,5} \varphi(F_i) = 14 \tag{16}$$

These numbers correspond respectively to the number of atoms for the prime numbers 5 (3×5) and 7 (2×7), see Table 3. Another way to compute the number of atoms in the second set $S^{II}$ is to evaluate it as a function of the number of NOS-atoms, which are Fibonacci numbers as shown above, in the respective sub-sets:

$$\sum_{i=1,2,3} [2(6-k) - F_i] + F_i = (9+1) + (7+1) + (4+2) = 24 \tag{17}$$

and

$$\sum_{i=4,5} (15-i) + F_i = (11+3) + (10+5) = 29 \tag{18}$$

In Eqs.(17) and (18), the first number in each parenthesis is the number of carbon atoms and the second one the number of NOS atoms. At this point, it is worth to re-examine the carbon atom content in the 20 amino acids. Some years ago, Yang (2003), identified the Lucas series (related to the Fibonacci series) in the great majority of the carbon atoms of the side-chains of the 20 amino acids (18 out of 20). The Lucas series $L_i$ are defined by the recurrence relation $L_i = L_{i-1} + L_{i-2}$ (i>1) with the initial conditions $L_0=2$, $L_1=1$. The first few terms are {2, 1, 3, 4, 7, 11, 18, …}. The carbon atom-numbers appearing in the side-chains of the 20 amino acids are 0, 1, 2, 3, 4, 7 and 9, and we see that, with the exception of 0 (Glycine) and 9 (Tryptophane), all are Lucas numbers (recall that, from Table 3, the number of carbon atoms for an amino acid is equal to the number of atoms minus the number of NOS atoms). There are also multiplicities, for example, there are 3 amino acids with 2 carbon atoms, 5 amino acids with 3 carbon atoms, and so on. Let us arrange the 20 amino acids, according to *increasing* multiplicity (second row)

| carbon atom-number: | 0 | 9 | 7 | 1 / 2 | 3 / 4 |
|---|---|---|---|---|---|
| multiplicity: | 1 | 1 | 2 | 3 | 5 |

Presented in this way, we have for the multiplicities a Fibonacci sequence (1, 1, 2, 3, 5) but in reality the numbers 3 and 5 appear two times, a multiplicity of multiplicity. If we separate each one of these repeating figures from the others and add the numbers, we get 1+1+2+3+5+(3+5) =1+1+2+3+5+8. This is the sum, 20, of the first six Fibonacci numbers and, of course, the number of amino acids. The product of these multiplicities by the corresponding numbers of carbon atoms gives (0×1+9×1+7×2+1×3+3×5) + (2×3+4×5)=41+26=67, the total carbon atom-number in the 20 amino acids. Note that 41 and 26 are the numbers of carbon atoms in the two sets $S^{II}$ and $S^{I}$, respectively, see more below. The above sum of the multiplicities, 1+1+2+3+5+(3+5), could also be written (1+1+2+3+5) + (1+1+1+2+2+1), where we have used the definition of the Fibonacci numbers to write 3 as 2+1→1+1+1 and 5 as 2+3→2+2+1. The last two results, taken together, lead to [41+(1+1+2+3+5)] + [26+(1+1+1+2+2+1)]=53+34. This is a very significant result as the two terms above describe nicely the C/NOS structure of the two sets $S^{II}$ and $S^{I}$ of Table 3, respectively: in $S^{I}$, there are 26 carbon atoms and 1+1+1+2+2+1=8 NOS atoms and, in $S^{II}$, there are 41 carbon atoms and 1+1+2+3+5=12 NOS atoms (Table 3 and Eqs.12-16). The above scheme, i.e., the carbon atom sequence and *two times* the multiplicity (Fibonacci) sequence, that we have used to arrive at this conclusion resembles the one of Table 1 in section 2 (the doubled Fibonacci sequence). *We have thus obtained the total number of atoms* (in $S^{I}+S^{II}$) *from the carbon atom content, only*. To end this section, let us show that the (missing) number of hydrogen atoms, as well as the number of nucleons, in the two sets $S^{I}$ and $S^{II}$ of Table 3, could be computed, using the significant Fibonacci number $F_9$= 34, only, and some few of its simple arithmetic functions. Amongst these latter, first, the $B_0$-function of an integer n (Négadi, 2009) which is the sum of the prime factors, the sum of their indices and the number of the prime factors of n, the so-called $\Omega$-function. As an example, take the number 34 itself. Its prime factorization is 2×17 and we have $B_0(34)$=(2+17)+(1+7)+2=29, because 2 is the 1st prime, 17 the 7th prime. Second, we use the well-known sum of the divisors function $\sigma(n)$. Still for our above example, 34 has four divisors 1, 2, 17 and 34, so that $\sigma(34)$=54. Now, we compute the hydrogen atom content in $S^{I}$ and $S^{II}$. Define the following expression

$$F_9+B_0(F_9) + \sigma(F_9) \tag{19}$$

The function $B_0$ has the "logarithmic" property $B_0(34)=B_0(2\times17)=B_0(2)+B_0(17)$ so that we could rewrite it as

$$[34+B_0(17)] + [B_0(2) + \sigma(F_9)]=[34+25]+[4+54]=59+58=117 \tag{20}$$

This is the total number of hydrogen atoms in the 20 amino acids (Négadi, 2009)[1], 59 in $S^{I}$ and 58 in $S^{II}$. We could also present the result in the two equivalent forms (34+25)+(7+51) or (34+25)+(24+34). Interestingly, we note that in $S^I$ there are 34 hydrogen atoms for the amino acids with no NOS atoms and 25 hydrogen atoms in the amino acids possessing NOS atoms. Also, in $S^{II}$ there are 7 hydrogen atoms in the sole amino acid Phe, having no NOS atoms, and 51 hydrogen atoms in the rest of amino acids all possessing NOS atoms. Note also that, in this case, we have 24 odd hydrogen numbers and 34 even hydrogen numbers (see footnote 1). We have therefore the correct number of HCNOS atoms in the two sets $S^I$ and $S^{II}$: 34+59=93 and 53+58=111, respectively, with total sum 204. The nucleon content in the 20 amino acids has been shown by several researchers studying the genetic code to be an important quantity, leading to many significant and beautiful mathematical patterns (Négadi, 2014; 2009 and the references therein). Here, we could compute, easily, the number of nucleons in the two sets $S^I$ and $S^{II}$ and, this time, we invoke another elementary function, $\tau(n)$, the number of divisors function. We have, with $F_9=34$, $B_0(F_9)=29$ and $\tau(F_9)=4$

$$[F_9 + 2B_0(F_9)] \times [B_0(F_9) + 2\tau(F_9)] = 2 \times 29^2 + [34(29+8) + 58 \times 8]$$
$$= 1682 + 1722 = 3404 \tag{21}$$

---

[1] Hydrogen atom-numbers and nucleon numbers of the amino acids side-chains, respectively, for the two sets $S^{I}$: {Gly (1, 1), Ala (3, 15), Ser (3, 31), Cys (3, 47), Pro (5, 41), Thr (5, 45), Val (7, 43), Asn (4, 58), Asp (3, 59), Ile (9, 57), Leu (9, 57), Met (7, 75)}; $S^{II}$: {Gln (6, 72), Glu (5, 73), Lys (10, 72), Arg (10, 100), Phe (7, 91), His (5, 81), Tyr (7, 107), Trp (8, 130)}, see Négadi, 2009.

This is the exact number of nucleons in the 61 amino acids (degeneracy included) partitioned into 1682 nucleons in $S^I$ and 1722 nucleons in $S^{II}$ (see footnote 1). Let us return, briefly, to Eqs.(6)-(7) of section 2 to say something concerning the distribution of the codons in the two sets $S^I$ and $S^{II}$. Looking at Table 3, we have 42 codons in $S^I$ (5 quartets, 2 sextets, 3 doublets, 1 triplet, 1 singlet) and 19 codons in $S^{II}$ (1 sextet, 6 doublets, 1 singlet). Now, from Eqs.(6) and (7), we have 10+19+10+22=61. It suffices to separate the three even numbers from the only odd number 19 to get 42+19=61. This is the codon-number pattern of $S^I$ and $S^{II}$, respectively. Let us make a final remark in this section still concerning the two sets $S^I$ and $S^{II}$. Consider, first, the number of amino acids in $S^I$, *in the detail*, and the number of those in $S^{II}$, *as a whole*. We have the sequence {1, 1, 2, 3, 5; 8}. Now, consider the NOS atom-numbers in $S^{II}$, *in the detail*, and those in $S^I$, *as a whole*. We have the *same* sequence {1, 1, 2, 3, 5; 8}. In the first case, the reading is top-down and, in the second case, bottom-up and we have a double sequence pattern, in a symmetrical (or anti-symmetrical, if one prefers) layout. We suspect that this property could somehow constitute an *a-posteriori* "justification" for the choice of the useful double Fibonacci sequence of Table 1, in section 2, which led us to establish the multiplet structure of the genetic code, its degeneracy as well as to several other interesting and significant results.

**4. Unifying the genetic code, and its variants, using q-Fibonacci numbers**
The concept of q-deformations, or "quantum deformations", in modern parlance, is quite ancient; it goes back to Jackson's q-calculus (Jackson, 1904) and even to Euler. In mathematics, it consists in the introduction of a parameter q in a mathematical function (number, binomial, trigonometric functions, special functions, and so on), or a theorem, in such a manner that, when q→1, the classical function, or theorem, is recovered. The simplest example (Euler) is the q-deformation of a natural number n:

$$[n]_q = \frac{1-q^n}{1-q} = 1 + q + q^2 + \cdots + q^{n-1} \tag{22}$$

For example $[3]_q=1+q+q^2 \to 3$ when $q \to 1$. In these last decades, quantum groups and quantum algebras, relying on the concept of q-deformations, have been introduced and largely used for applications, mainly in physics and also in chemistry. For example, such important subjects as the harmonic oscillator (Macfarlane, 1989) the hydrogen atom (Kibler and Négadi, 1991), or the "Aufbau Prinzip", which is at the root of the study of the periodic table of the elements (Kibler and Négadi, 1992), have been considered. The reader could also find, in the latter three cited papers, many references concerning the applications (solid-state physics, statistical physics, nuclear physics, and so on). It appears that some subtle effects, for example, in the vibratory and rotatory complex structure of molecules and atomic nuclei coud be taken into account, efficiently, when using q-deformed models. In this section, we shall consider the q-Fibonacci numbers, as our q-deformed mathematical objects. At this stage, let us make an important remark. In physics, the objects to be q-deformed exist and are determined from the existing mathematical equations as, for example, the harmonic oscillator, the Schrödinger equation, the entropy, etc.. In the case of the genetic code, there is no such existing fundamental equations. For this reason, in this work, we decided to take as q-deformed objects the Fibonacci numbers, considered in the second section, inasmuch as they are related to the number of amino acids and to the degeneracy of the genetic code and also the fact that the q-Fibonacci numbers exist. These latter were introduced, in mathematics, by Carlitz forty years ago (Carlitz, 1974). They obey the following recurrence relation

$$f^q_{n+1} = f^q_n + q^{n-1} f^q_{n-1} \tag{23}$$

where q is the deformation parameter (assumed here to be a positive real number) and we have that $f_{n+1}=f_n+f_{n-1}$ when q=1, i.e., the usual recurrence relation for the ordinary Fibonacci numbers. The first few q-Fibonacci numbers are given below (Carlitz, 1974)

$$f^q_0 = 0; \; f^q_1 = 1 \tag{24}$$

$$\begin{aligned} f^q_2 &= 1 \\ f^q_3 &= 1 + q \\ f^q_4 &= 1 + q + q^2 \\ f^q_5 &= 1 + q + q^2 + q^3 + q^4 \\ f^q_6 &= 1 + q + q^2 + q^3 + 2q^4 + q^5 + q^6 \end{aligned} \tag{25}$$

In the limit q=1, it is easily seen that these expressions reduce to the usual Fibonacci numbers (see the introduction). First, invoking again the double sequence of Table 1, we consider as q-deformed objects the Fibonacci numbers of the sequence $f_i$, *only*, and define the sum

$$\lfloor \sum_{i=1,2,\ldots,6} i + f^q_i \rfloor + \sum_{i=1,2,\ldots,6} F_i \tag{26}$$

By replacing the functions $f_i{}^q$, from Eqs.(24)-(25), in the above relation, we get

$$= \lfloor 27 + 4q + 3q^2 + 2q^3 + 3q^4 + q^5 + q^6 \rfloor + 20 \qquad (27)$$

In these expressions, $\lfloor x \rfloor$ is the mathematical "floor" function of x (see section 3 for the definition). This is to ensure that the result is an *integer*, as the degeneracies (number of codons) are discrete quantities (integers). Recall that the Fibonacci numbers of the sequence $f_i$ and their indices are connected to the degeneracy, see section 2. For q=1, we recover the result in Eq.(5) of section 2 for the standard genetic code, with 20 amino acids and 41 degenerate codons. For q≠1, the results are different and, for q~1, we are going to show below that, in this case, we could describe the degeneracy structure of the various variants of the standard genetic code[2] and more. First, let us start by considering these variant genetic codes which could be grouped into four categories, according to their number of stop codons because the number of degenerate codons (*globally* the same in each category) is equal to 64 (the total number of codons) minus 20, the number of amino acids, which is the same for all the variant genetic codes, minus the number of stop codons. Therefore, the number of these latter characterizes the four categories which are (i) 4 stop codons: the *Vertebrate Mitochondrial Code*, (ii) 3 stop codons: the *Standard Code*, the *Bacterial, Archaeal and Plant Plastid Code*, the *Alternative Yeast Nuclear Code*, *Scenedesmus obliquus Mitochondrial Code*, *Thraustochytrium Mitochondrial Code*, (iii) 2 stop codons: the *Yeast Mitochondrial Code*, the *Mold, Protozoan*, and *Coelenterate Mitochondrial Code* and the *Mycoplasma/Spiroplasma Code*, the *Invertebrate Mitochondrial Code*, the *Echinoderm* and *Flatworm Mitochondrial Code*, the *Euplotid Nuclear Code*, the *Ascidian Mitochondrial Code*, *Chlorophycean Mitochondrial Code*, *Trematode Mitochondrial Code*, *Pterobranchia Mitochondrial Code*, *Candidate Division SR1* and *Gracilibacteria Code*, and finally (iv) 1 stop codon: the *Ciliate, Dasycladacean* and *Hexamita Nuclear Code*, the *Alternative Flatworm Mitochondrial Code*. Using Eqs.(27), it appears that these four cases could be described by the following q-values:

$$\text{(i)} \quad q \sim 0.974: \quad (21 + 19) + 20 = 40 + 20 = 60 \qquad (28)$$

$$\text{(ii)} \quad q = 1: \quad (21 + 20) + 20 = 41 + 20 = 61 \qquad (29)$$

$$\text{(iii)} \quad q \sim 1.025: \quad (21 + 21) + 20 = 42 + 20 = 62 \qquad (30)$$

$$\text{(iv)} \quad q \sim 1.049: \quad (21 + 22) + 20 = 43 + 20 = 63 \qquad (31)$$

In case (i), we have 20 amino acids and 40 degenerate codons, in case (ii), 20 amino acids and 41 degenerate codons, in case (iii), 20 amino acids and 42 degenerate codons and finally, in case (iv), 20 amino acids and 43 degenerate codons. Here also, as mentioned in the remark following Eq.(4) of section 2, we could subtract each one of the equations (28)-(31) from 64 to get the corresponding number of stop codons 4, 3, 2 and 1, respectively. For example, in case (i), we have 64-(20+40)=4 (or 40+20+4=64). Second, we consider the case of additional amino acids. Selenocysteine (Sec) and Pyrrolysine (Pyl) are rare amino acids which have been discovered in 1986 (Böck *et al.*, 1991) and 2002 (Srinivasan *et al.*, 2002), respectively, and have joined the "exclusive club" of the twenty canonical amino acids; they are also known as the 21$^{st}$ and 22$^{nd}$ amino acids, respectively. Both are cotranslationally encoded by the codons UGA and UAG, respectively, these latter function usually as stop codons. Here, the quantity which will be deformed is the number of amino acids, that is the numbers of the other Fibonacci sequence, $F_i$, see section 2, and the sequence $f_i$ remains unaltered, as the new amino acids are coded by the stop codons so that the overall degeneracy is unaltered. We write therefore, using again Eqs.(24)-(25)

$$\sum_{i=1,2,...,6} i + f_i + \lfloor \sum_{i=1,2,...,6} F_i \rfloor \tag{32}$$

$$= 41 + \lfloor 6 + 4q + 3q^2 + 2q^3 + 3q^4 + q^5 + q^6 \rfloor$$

---

[2] http://www.ncbi.nlm.nih.gov/Taxonomy/Utils/wprintgc.cgi

First, the case where only one of these additional amino acids is encoded, at a time, is the same as in case (iii) above with $f_i \to F_i$, so that, for q~1.025, we get 41+21=62, 41 degenerate codons and 21 amino acids (the 20 canonical ones and one more, either Selenocysteine or Pyrrolysine). Here the degeneracy, which is related to the sequence $f_i$, remains unaltered, as mentioned above, and the change is only in the amino acids number. The number of stops is here equal to 2 (64-62). It is interesting to note that the case where Selenocysteine and Pyrrolysine are *both* encoded has been shown to exist (Zhang and Gladyshev, 2007). In the latter reference, the authors have shown the existence of large selenoproteomes and pyrroproteomes in symbiotic deltaproteobacteria of a gutless worm *Olavius Algarvensis*. In this case, that is for two more encoded amino acids, the value of the deformation parameter would be q~1.049, as in case (iv) above, but using Eq.(32): 41+22=63, i.e., 41 degenerate codons and 22 amino acids (total 63 codons). In this case, we have only one stop left (64-63). Finally, we consider the interesting limiting case q=0. In this case, the Fibonacci sequence "degenerates" into the trivial sequence {1, 1, 1, 1, 1, 1, ...} and the sum of the first six terms, which gives the number of amino acids 20, in the case q=1, gives here 6, a rather small but nonetheless interesting number. As a matter of fact, it agrees with claims, old and still persistent today, that life, at its origin, could have started using only a small number of amino acids (or groups of amino acids). After the all first seminal speculations about this idea, made by Woese (1965) and Crick (1968) came Wong's Coevolution Theory (1965) which postulates that, quoting Wong, "The structure of the codon system is primarily an imprint of the prebiotic pathways of amino acids formation, which remain recognizable in the enzymic pathways of amino acid biosynthesis". He also writes in his paper "Accordingly, at the very early stages, the system of amino acids entering into primordial proteins likely consisted of only a few amino acids, each occupying a continuous domain of the genetic code". More recently, Di Giulio (Di Giulio, 2008) extended this theory and, interestingly, proposes that only six amino acids were involved in the early phases of the genetic code, just the right number we found above when the deformation parameter is equal to zero. He writes, in the abstract of his paper, "It is re-observed that the first amino acids to evolve along these biosynthetic pathways are predominantly those codified by codons of the type GNN, and this observation is found to be statistically significant. Furthermore, the close biosynthetic relationships between the sibling amino acids Ala-Ser, Ser-Gly, Asp-Glu, and Ala-Val are not random in the genetic code table and reinforce the hypothesis that the biosynthetic relationships between these six amino acids played a crucial role in defining the very earliest phases of the genetic code origin". Interestingly, there is a second line of research which proposes also six, as the number of first amino acids, it is the Trifonov-Bettecken GCU-based Theory (Trifonov and Bettecken, 1997; Trifonov, 2002). In his paper (Trifonov, 2002), he explains "Hidden periodical $(GCU)_n$ pattern in extant mRNA, and predominant $(GCN)_n$ repeat in the triplet expansion diseases suggest that the GCU triplet and its 9 point change derivatives could have been the first codons. The earliest six amino acids (A, D, G, P, S and T) are suggested by their chemical simplicity, by experiments of Stanley Miller, and by association with more ancient aminoacyl tRNA synthetases of class II. Strickingly, all these amino acids, indeed, are encoded by the above mentioned codons". Of course, our above proposed q-deformed model (with q=0), even if it fits nicely the conclusions of the extended Di Giulio Coevolution Theory, and also the Trifonov-Bettecken GCU-based Theory, remains only a toy model, as we have no information about the primordial degeneracies and the primordial number of stop codons. What we could eventually write, using Eq.(32), would be only the difference $64 - \lfloor 6+4q+3q^2+2q^3+3q^4+q^5+q^6 \rfloor$ giving the number of degenerate-and-stop codons and decreasing from 58 (for q=0), to 44 (for q=1) while, at the same time, the number of amino acids increases from 6 to 20 and even to 21 and 22 if one considers the additional amino acids Selenocysteine and Pyrrolysine (see Figure 1).

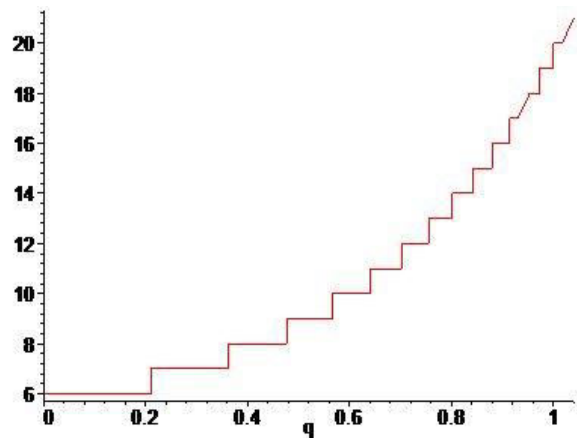

**Figure 1.** Evolution of the number of amino acids $\lfloor 6+4q+3q^2+2q^3+3q^4+q^5+q^6 \rfloor$ as a function of the deformation parameter q.

## 5. Concluding remarks

We have thus devoted this paper to show, we hope convincingly, that the first terms of the Fibonacci sequence, not only are well suited for a description of the genetic code mathematical structure (number of codons, degeneracy, characteristic patterns), but they also show themselves in a classification of the twenty canonical amino acids, sorted according to their increasing non-hydrogen atom numbers. They could, as well, serve to compute the detailed chemical composition in sub-sets of the above mentioned classification. Finally, the extension to their q-deformed version led us to establish a unified phenomenological scheme, depending on a deformation parameter q, describing, at the same time, the standard genetic code, its several non-standard variants, additional amino acids, such as Selenocysteine and Pyrrolysine, as well as a special case of a small number amino acids, maybe suited for early life scenario. *We stress that the Fibonacci sequence, used in this paper, to obtain good agreements or good fits, just as many other mathematical model functions lead to good fits, **is no explanation** of the underlying (complex) biological phenomena. As explained by Mäkelä and Annila in their reiew (Mäkelä and Annila 2010): "Universal patterns such as* *power-law dependences, skewed distributions, tree-like structures, networks and spirals are associated with energy dispersal processes using the principle of least action." The genetic code and the amino acids chemical composition have followed this principle and, just like the branched phylogenetic tree, the spiraled Nautilus and so on have been modeled by number sequences, they have been modeled, in this work, by the Fibonacci sequence.* We now end our paper with some remarks. First, in section 2, we seized an opportunity and began by presenting a new mathematical formula which reproduces, exactly, the multiplet structure of the (standard) genetic code: 6 sextets, 5 quartets, 9 doublets, 1 triplet and 2 singlets. It appears to us that, using the same mathematical ingredients, we could extend our result to include the case of the twenty or so known slight variations of the genetic code, the so-called non-standard genetic codes, thanks to the introduction of additional Kronecker delta functions (and/or simple shifts), playing the role of "perturbations". This seems to be sufficient to derive the multiplet structure, group-numbers and class-numbers, of the above mentioned genetic codes. As an illustration example, we consider, here, the case of the Vertebrate Mitochondrial Code. The group-numbers are given by $2^k+2(k-1)$ and the class-numbers by $8-2k$, as for the even class-numbers of the standard genetic code (k=1, 2, 3). Here, no additional Kronecker delta function is needed, a simple index-dependent shift in the group-number function is enough. It is readily seen, from the above two functions, that there are 2 sextets (k=1), 6 quartets (k=2) and 12 doublets (k=3), as it must be for this case (see footnote 2). We plan to publish new results for the remaining other cases in the near future (Négadi, 2015). Our second remark concerns the classification which was at the basis of our treatment in section 3. We have already mentioned that our Table 3 was inspired from a list of the 20 amino acids ranked according to their atom number given by Trifonov, in a presentation (Trifonov, 2001), where the author's intention was to show that the first members of his list are predominantly those amino acids found in the Urey-Miller experiments (indicated by a star in our Table 3). This list constitute the first one from forty criteria, used by Trifonov, to build a chronological order of appearance of amino acids in the early evolution of life. He called this first criterion, based on the ranking of the amino acids according to their increasing (non-hydrogen) atom numbers, the criterion of "simplicity" (Trifonov, 2000). He also divided his list into two separate parts, intentionally or not, we do not know, but this partition appears to us very interesting; these two parts are our $S^I$ and $S^{II}$ sub-sets, in Table 3. Interestingly, Higgs and Pudritz (2009), on the basis of a thermodynamic approach, found that the ten amino acids Gly, Ala, Asp, Glu, Val, Ser, Ile, Leu, Pro and Thr could constitute what they call the "early group" and their results are close to those of Trifonov, in Trifonov (2004) and, with the exception of Glu, all the above amino acids are in our sub-set $S^I$ in Table 3. In summary and, with some very few exceptions, the sub-set $S^I$ contains the "earliest" amino acids and $S^{II}$ contains the "latest". A further mathematical partition of $S^{II}$, into two other sub-sets, led us finally to the interesting results of section 3. The "simplicity" of this classification does not mean "triviality"; on the contrary, it means mathematical "beauty", as it shows several times the first members of the Fibonacci sequence and contains, in itself, the mathematical ingredients to compute its own detailed chemical composition (see section 3). Also, concerning the above mentioned few exceptions, the case of Cysteine and Methionine which belong to our set $S^I$, deserves a brief comment. These two sulfur-containing amino acids are generally not considered as early amino acids (Trifonov, 2000; 2004; Higgs and Pudritz, 2009). However, recent experiments by Bada and co-workers (Parker *et al.*, 2011) inspired from Miller's 1858 $H_2S$-experiment, have found, in their samples, among others, oxidation products of Cysteine and Methionine. When one thinks that oxygen was quasi-absent, if not absent, in the early atmosphere, then these two amino acids could have been present (in volcanic plumes, for example) and therefore could have been members of the early amino acids set but, of course, only further research will arbitrate. The third remark is about the various q-values, given in section 4 and accompanying the equations (28), (30) and (31). These values, q=0.974, q=1.025 and q=1.049, have been used in the latter equations to get exact integer results, using the "floor" function, as we have mentioned, there. Had we decided not to use the "floor" function, then the following respective fine-tuned values, 0.973439188, 1.024803313, and 1.048072264, would give "integer" results, with high precision. Note also that, in this case, the above q-values with three significant digits give, too, "integer" results with reasonable precision. Our last remark concerns the last step in our applications, in section 4, using the q-Fibonacci numbers. We have seen that the value of the deformation parameter q=1 describes the standard genetic code, that different q-values, near 1, are suited for the description of the non-standard versions of the genetic code as well for the inclusion of additional amino acids (Selenocysteine or/and Pyrrolysine) and, finally, that the (special) value q=0 gives the small amino acids number 6. We have then mentioned that this last number agrees well with proposals in existing works as the extended Di Giulio Coevolution and Trifonov-Bettecken GCU-based theories. We have assumed, in section 4, that the deformation parameter q is a positive real number but, in fact, nothing prevents us to expand it to negative values inasmuch as the function describing the number of amino acids, $6+4q+3q^2+2q^3+3q^4+q^5+q^6$ (see Eq.(32)), has a minimum for q<0. Using a Maple solving package, we find that the value q~ −0.613, where the derivative of the above function (a quintic equation) vanishes, corresponds to the minimum and the function takes, here, the value 4.6 or 4 by taking its "floor" function, as we have done in our equations in section 4. This latter value, 4, is smaller than the one obtained for q=0 and it corresponds to a minimum for the number of amino acids. Strikingly, the above value of the deformation parameter, −0.613, leading to a minimum, is very close to the golden mean φ=−0.618… (including sign!). Recall that the golden mean (or golden section), which is intimately connected to the Fibonacci sequence, is also defined by the equation $x^2-x-1=0$ and the two solutions of this equation are Φ=1.618… and φ=−0.618…. In this new situation, having a-priori no good reason to discard the negative values for the deformation parameter q, we must include, also, other recent interesting works, proposing a number of "earliest" amino acids less than six but certainly greater than or equal to four  (Higgs, 2009; Hartman and Smith, 2014).